\documentclass[twocolumn,preprintnumbers,amsmath,amssymbm,prd]{revtex4}
\usepackage{epsfig}
\usepackage{graphicx}

\begin{document}
\title{Quantum buoyancy, generalized second law, and higher-dimensional entropy bounds}
\author{Shahar Hod}
\affiliation{The Ruppin Academic Center, Emeq Hefer 40250, Israel}
\affiliation{ } \affiliation{The Hadassah Institute, Jerusalem
91010, Israel}
\date{\today}

\begin{abstract}
\ \ \ Bekenstein has presented evidence for the existence of a
universal upper bound of magnitude $2\pi R/\hbar c$ to the
entropy-to-energy ratio $S/E$ of an arbitrary {\it three}
dimensional system of proper radius $R$ and negligible self-gravity.
In this paper we derive a generalized upper bound on the
entropy-to-energy ratio of a $(D+1)$-dimensional system. We consider
a box full of entropy lowered towards and then dropped into a
$(D+1)$-dimensional black hole in equilibrium with thermal
radiation. In the canonical case of three spatial dimensions, it was
previously established that due to quantum buoyancy effects the box
floats at some neutral point very close to the horizon. We find here
that the significance of quantum buoyancy increases dramatically
with the number $D$ of spatial dimensions. In particular, we find
that the neutral (floating) point of the box lies near the horizon
only if its length $b$ is large enough such that $b/b_C>F(D)$, where
$b_C$ is the Compton length of the body and $F(D)\sim D^{D/2}\gg1$
for $D\gg1$. A consequence is that quantum buoyancy severely
restricts our ability to deduce the universal entropy bound from the
generalized second law of thermodynamics in higher-dimensional
spacetimes with $D\gg1$. Nevertheless, we find that the universal
entropy bound is always a sufficient condition for operation of the
generalized second law in this type of gedanken experiments.
\end{abstract}
\bigskip
\maketitle


\section{Introduction}
One of the most intriguing features of both the classical and
quantum theory of black holes is the striking analogy between the
laws of black-hole physics and the universal laws of thermodynamics
\cite{Chris,Haw1,BarCar}. In particular, Hawking's (classical)
theorem \cite{Haw1}, ``The surface area of a black hole never
decreases", is a property reminiscent of the entropy of a closed
system. This remarkable analogy had led Bekenstein \cite{Bek2} to
conjecture that the area of a black hole (in suitable units) may be
regarded as the black-hole entropy -- entropy in the sense of
inaccessible information about the internal state of the black hole.
This conjecture is logically related to a second conjecture, known
as the generalized second law of thermodynamics (GSL) \cite{Bek2}:
``The sum of the black-hole entropy ($1/4$ of the horizon's surface
area) and the common (ordinary) entropy in the black-hole exterior
never decreases."

Arguing from the GSL, Bekenstein \cite{Bek3} has proposed the
existence of a universal upper bound on the entropy $S$ (and hence
information capacity) of any weakly self-gravitating physical system
of circumscribing radius $R$ and total energy $E$:
\begin{equation}\label{Eq1}
S\leq {{2\pi RE}\over{\hbar c}}\  .
\end{equation}
This bound was deduced from the requirement that the GSL be
respected when a box containing entropy is placed with no radial
motion near the horizon of a black hole, and then allowed to fall
in. The entropy of the box disappears but an increase in the black
hole entropy occurs. The GSL is respected provided the box's entropy
$S$ is bounded as in (\ref{Eq1}).

It is worth mentioning that, the canonical Bekenstein bound
(\ref{Eq1}) can be tightened: an improved upper bound to the entropy
of a spinning object was obtained in \cite{Hodspin} and an improved
bound to the entropy of a charged system was found in
\cite{BekMay,Hodcharge}. For a cubic system, a somewhat stronger
bound can be deduced from the GSL with $2R$ replaced by the cubic
length $b$:
\begin{equation}\label{Eq2}
S\leq {{\pi bE}\over{\hbar c}}\  .
\end{equation}

The main purpose of the present paper is to extend Bekenstein's
analysis to {\it higher} dimensional spacetimes. We shall find below
that in the generic case of $D$ flat spatial dimensions, the bound
(\ref{Eq2}) may be deduced from the GSL only under severe
restrictions on the size of the system.

\section{Acceleration radiation}
In the original derivation \cite{Bek3} of the universal entropy
bound it was assumed that the energy (as measured at infinity) added
to the black hole with the box is that which may be inferred from
the red-shift factor at the deposition point (this energy determines
the increase in black-hole entropy). However, it was pointed out in
\cite{UW} that when the deposition of the box is attempted by
lowering it from far away, buoyancy in the radiative black hole
environs will prevent lowering the box all the way down to the
horizon \cite{Bek5}. In fact, it was realized \cite{UW} that one
must invest energy in order to push the box into the black hole
against the quantum buoyant force \cite{Bek5}. This extra energy
contributes to the black-hole area increase, helping to ensure the
validity of the GSL in this gedanken experiment.

An observer accelerating with acceleration $a$ detects isotropic
thermal radiation with temperature
\begin{equation}\label{Eq3}
T_U={{\hbar a}\over{2\pi k_Bc}}\  ,
\end{equation}
the celebrated Unruh radiance \cite{UW}. An object hanging above a
black hole is accelerated by virtue of its being prevented from
following a geodesic \cite{Bek4}. It was suggested \cite{UW} that
this object will likewise see Unruh radiance. Because its
acceleration (and hence the temperature $T_U$) varies with the
height above the horizon, it was concluded that the object will be
influenced by a quantum buoyant force. This buoyant force is due to
the non-uniformity of the ambient pressure (non-uniformity of the
Unruh temperature) \cite{Bek4}.

Two intriguing consequences were inferred \cite{UW,Bek5,Bek4}:
\begin{itemize}
\item {The buoyancy can cause an object sufficiently near the horizon to
float at some neutral point.}
\item{The buoyancy contributes to the energy bookkeeping of a process in which a box is lowered from
far away towards a black hole while doing work on the lowering
machine. In particular, the total energy at infinity added to the
hole after the box has been dropped from the neutral point is {\it
larger} than the red-shifted proper energy of the box.}
\end{itemize}

It should be emphasized that it was shown in \cite{Bek5} that for
generic systems in {\it three} spatial dimensions the neutral point
lies very near the horizon. A consequence is that in three spatial
dimensions, the buoyancy makes only a negligible contribution to the
energy bookkeeping of the gedanken experiment. This implies that the
original entropy bounds (\ref{Eq1})-(\ref{Eq2}) can be deduced if
the GSL is assumed to hold. However, in this paper we shall show
that this conclusion may not be valid in generic $(D+1)$-dimensional
spacetimes with $D\gg 1$.

We shall carry out the gedanken experiment with a
$(D+1)$-dimensional Schwarzschild-Tangherlini black hole
\cite{SchTang,Kun} of ADM mass ${\cal M}$. The exterior spacetime is
described by the metric (we use natural units in which $G=c=k_B=1$)
\begin{equation}\label{Eq4}
ds^2=-H(r)dt^2+{H(r)}^{-1}dr^2+r^2d\Omega^{(D-1)}\ ,
\end{equation}
where
\begin{equation}\label{Eq5}
H(r)=1-{\Big({r_H\over r}\Big)}^{D-2}\equiv [\chi(r)]^2
\end{equation}
defines the redshift factor, $\chi(r)$. Here
\begin{equation}\label{Eq6}
r_H={\Big[{{16\pi {\cal M}}\over{(D-1)A_{D-1}}}\Big]}^{1\over{D-2}}
\end{equation}
is the black hole's radius and
\begin{equation}\label{Eq7}
A_{D-1}={{2\pi^{D/2}}\over {\Gamma(D/2)}}
\end{equation}
is the area of a unit $(D-1)$-sphere.

Following \cite{UW,Bek5,Bek4} we assume the black hole has reached
equilibrium with its own Hawking radiation, the whole system being
enclosed in a large cavity. The black hole temperature $T_{bh}$ and
the local temperature $T$ are related by \cite{UW,Bek5,Bek4}
\begin{equation}\label{Eq8}
T={{T_{bh}}\over{\chi}}={{(D-2)\hbar}\over{4\pi r_H \chi}}\  .
\end{equation}
In the
near-horizon region, $r-r_H\ll r_H$, the redshift factor is given by
\begin{equation}\label{Eq9}
\chi^2(r)=(D-2){{r-r_H}\over{r_H}}[1+O((r-r_H)/r_H)]\  .
\end{equation}
The radial coordinate $r$ is related to the proper distance $l$
above the horizon through the relation
\begin{eqnarray}\label{Eq10}
&&l(r)=\int_{r_H}^{r}[\chi(r')]^{-1}dr' \nonumber \\&&
=2\Big[{{r_H(r-r_H)}\over{D-2}}\Big]^{1/2}[1+O((r-r_H)/{r_H})]\ .
\end{eqnarray}
From (\ref{Eq9}) and (\ref{Eq10}) one obtains the near-horizon
relation
\begin{equation}\label{Eq11}
\chi(l)={{D-2}\over{2r_H}}l\  .
\end{equation}
Taking cognizance of Eq. (\ref{Eq8}), one can write the local
temperature as
\begin{equation}\label{Eq12}
T={{\hbar}\over{2\pi l}}\  .
\end{equation}

Note that one could also obtain (\ref{Eq12}) from the observation
\cite{Bek5,Bek4} that a point suspended at rest in a
$(D+1)$-dimensional Schwarzschild spacetime is characterized by an
invariant acceleration $a=d\chi/dr$. Taking cognizance of Eqs.
(\ref{Eq3}) and (\ref{Eq11}), one obtains the formal Unruh
temperature $\hbar/2\pi l$.

\section{Energetics of the system}
The main goal of this paper is to extend Bekenstein's analysis to
higher dimensional spacetimes, properly taking into account the
contribution of quantum buoyancy to the energy bookkeeping of the
lowering process. {\it Ignoring} the effects of quantum buoyancy, it
was found in \cite{Bous1} that the bound (\ref{Eq1}) can be inferred
from the GSL in $(D+1)$-dimensional spacetimes for any value of $D$.
However, we shall find below that the contribution of quantum
buoyancy to the energy bookkeeping increases dramatically with the
number of spatial dimensions $D$. As a consequence, we shall show
that the argument justifying the bound (\ref{Eq1}) from the GSL
cannot be completed for $D\gg 1$ unless the size of the object is
very large [see Eqs. (\ref{Eq34})-(\ref{Eq35}) below].

The system we consider is a $(D+1)$-dimensional hypercube with sides
of length $b$, rest mass $\mu$, and which holds entropy $S$. To
obtain an entropy bound, one may lower the system from infinity into
a $(D+1)$-dimensional black hole. The test-particle approximation
imposes the constraint $r_H(\mu)\ll b\ll r_H({\cal M})$. This
guarantees that the object has a negligible self-gravity and that it
is much smaller than the size of the black hole. One would like to
add as little energy as possible to the black hole. This will
minimize the corresponding increase of the black hole's surface area
and thus will help to optimize the tightness of the entropy bound
\cite{Bek3,Bous1}. Therefore, the strategy is to extract work from
the system by lowering the box slowly towards the black hole, before
one finally drops it in.

The total energy ${\cal E}$ (energy-at-infinity) of a body located
at a radial coordinate $r$ in the black-hole spacetime is made up of
two contributions \cite{Bek5,Bek4}:
\begin{itemize}
\item {${\cal E}_0 = \mu\chi(r)$, the energy associated
with the body's mass ({\it red-shifted} by the gravitational field).
Taking cognizance of Eq. (\ref{Eq11}), one finds
\begin{equation}\label{Eq13}
{\cal E}_0={{D-2}\over{2r_H}}\mu l\  .
\end{equation}}
\item {The contribution to the energy from the work done to
overcome the buoyancy is \cite{Bek5,Bek4}
\begin{equation}\label{Eq14}
W_{\text{buoy}}=\int_{l}^{\infty}f_{\text{buoy}}dl'\  ,
\end{equation}
where $l$ is the proper height of the centroid plane of the box
above the horizon, and $f_{\text{buoy}}$ is the buoyant force acting
on the box, as measured by an observer at infinity.}
\end{itemize}

The buoyant force acting on the box, as measured by an observer at
infinity, is the difference of the redshifted local forces acting on
the upper and lower faces \cite{UW,Bek4,Bek5}:
\begin{equation}\label{Eq15}
f_{\text{buoy}}(l)=A[(P\chi)_{l-b/2}-(P\chi)_{l+b/2}]\  ,
\end{equation}
where $A=b^{D-1}$ is the horizontal crossectional area of the box
and $P$ is the radiation pressure. As a consequence of the
cancelation of the work done by buoyant forces on top and bottom of
the box over the range $[l+b/2,\infty]$, the buoyant contribution to
the energy ${\cal E}$ of the body only depends on the distribution
of $P\chi$ over the height of the box. Taking cognizance of
(\ref{Eq15}) one therefore finds \cite{UW,Bek4,Bek5}
\begin{equation}\label{Eq16}
W_{\text{buoy}}=A\int_{l-b/2}^{l+b/2} P\chi dl^{'}\  .
\end{equation}

Following \cite{UW,Bek4,Bek5} we shall assume a model of
acceleration radiation as a mixture of noninteracting thermal gases
of massless particles. The mean energy density of thermal radiation
in $D$ spatial dimensions for one helicity degree of freedom is
given by \cite{Bek5}
\begin{equation}\label{Eq17}
e={\int_0^{\infty}}{{\hbar\omega\
dV_D(\omega)}\over{{(e^{\hbar\omega/T}\mp 1)}{(2\pi)^D}}}\  ,
\end{equation}
where the upper (lower) signs correspond to boson (fermion) fields,
and
\begin{equation}\label{Eq18}
dV_D(\omega)=[2\pi^{D/2}/\Gamma(D/2)]\omega^{D-1}d\omega\
\end{equation}
is the volume in frequency space of the shell
$(\omega,\omega+d\omega)$.

From Eqs. (\ref{Eq17})-(\ref{Eq18}) and the relation
\begin{equation}\label{Eq19}
{\int_0^{\infty}}{{x^Ddx}\over{e^x\mp1}}=\zeta(D+1)\Gamma(D+1)\times
\begin{cases}
1 & \text{for bosons} \ ; \\
1-2^{-D} & \text{for fermions} \ ,
\end{cases}
\end{equation}
where $\zeta(z)$ is the Riemann zeta function, one finds that the
mean energy density of all massless fields is given by
\begin{equation}\label{Eq20}
e={{ND\zeta(D+1)\Gamma({{D+1}\over{2}})T^{D+1}}\over{{\pi}^{{D+1}\over{2}}}\hbar^D}\
,
\end{equation}
where $N$ is the effective number of massless degrees of freedom
(the number of polarization states). Massless scalars contribute $1$
to $N$, massless fermions contribute $1-2^{-D}$ to $N$ \cite{Bek5},
an electromagnetic field contributes $D-1$ to $N$ \cite{CarCavHal},
and the graviton contributes $(D+1)(D-2)/2$ to $N$ \cite{CarCavHal}.
The thermal radiation pressure in $D$ spatial dimensions is given by
$P=e/D$ \cite{Aln}, yielding
\begin{equation}\label{Eq21}
P=e/D={{N\zeta(D+1)\Gamma({{D+1}\over{2}})T^{D+1}}\over{{\pi}^{{D+1}\over{2}}}\hbar^D}\
,
\end{equation}

Taking cognizance of Eqs. (\ref{Eq11}), (\ref{Eq12}), and
(\ref{Eq21}), one can write Eq. (\ref{Eq16}) as
\begin{equation}\label{Eq22}
W_{\text{buoy}}=A\int_{l-b/2}^{l+b/2}
{{N\zeta(D+1)\Gamma({{D+1}\over{2}})(D-2)\hbar}\over{2^{D+2}{\pi}^{{3D+3}\over{2}}r_Hl^{'D}}}dl^{'}\
.
\end{equation}
Performing the integration in (\ref{Eq22}), one finds
\begin{eqnarray}\label{Eq23}
&&W_{\text{buoy}}={{NA\zeta(D+1)\Gamma({{D+1}\over{2}})(D-2)\hbar}
\over{2^{D+2}\pi^{{3D+3}\over{2}}(D-1)r_H}}\nonumber \\&&
\times\Big[(l-{b\over 2})^{-D+1}-(l+{b\over 2})^{-D+1}\Big]\
\end{eqnarray}
for the energy contribution due to the work done to overcome the
quantum buoyancy. Finally, putting together Eqs. (\ref{Eq13}) and
(\ref{Eq23}), one finds
\begin{eqnarray}\label{Eq24}
&&{\cal E}(l)={{D-2}\over{2r_H}}\mu
l+{{NA\zeta(D+1)\Gamma({{D+1}\over{2}})(D-2)\hbar}
\over{2^{D+2}\pi^{{3D+3}\over{2}}(D-1)r_H}}\nonumber \\&&
\times\Big[(l-{b\over 2})^{-D+1}-(l+{b\over 2})^{-D+1}\Big]\
\end{eqnarray}
for the total energy of a body suspended at a proper distance $l$
above the horizon.

\section{The neutral point}
The most challenging test of the GSL is obtained by dropping the
object from the neutral (floating) point where ${\cal E}(l)$ reaches
its minimum \cite{UW,Bek4,Bek5}. Setting $d{\cal E}/dl=0$ in
(\ref{Eq24}), one obtains the condition determining the proper
distance $l_0$ of the floating point from the horizon:
\begin{eqnarray}\label{Eq25}
(l_0-{b\over 2})^{-D}-(l_0+{b\over
2})^{-D}={{2^{D+1}\pi^{{3D+3}\over{2}}\mu}\over{NA\zeta(D+1)\Gamma({{D+1}\over{2}})\hbar}}.
\end{eqnarray}
Substituting (\ref{Eq25}) back into (\ref{Eq24}), one finds
\begin{eqnarray}\label{Eq26}
&&{\cal E}_{\text{min}}={{D-2}\over{2r_H}}\mu \nonumber \\&& \times
\Bigg[l_0+{{1}\over{D-1}}{{(l_0-{b\over 2})^{-D+1}-(l_0+{b\over
2})^{-D+1}}\over{(l_0-{b\over 2})^{-D}-(l_0+{b\over 2})^{-D}}}\Bigg]
\end{eqnarray}
for the energy of the body at the floating point.

The universal bound on entropy, Eq. (\ref{Eq2}), can be deduced from
the GSL only when the optimal drop point is {\it close} to the
horizon, $l_0-b/2\ll b$. It is only in such situations that the
contribution of quantum buoyancy to the energy of the body [the
second term in Eq. (\ref{Eq26})] is small.

Under which conditions does the neutral point lie in the
near-horizon region? Substituting $l_0=b/2+\epsilon b$ with
$\epsilon\ll 1$ into Eqs. (\ref{Eq25})-(\ref{Eq26}), one finds
\begin{eqnarray}\label{Eq27}
&&{\cal E}_{\text{min}}={{D-2}\over{4r_H}}\mu
b\Big[1+{{2D}\over{D-1}}\epsilon+O(\epsilon^2)\Big]\  ,
\end{eqnarray}
with
\begin{eqnarray}\label{Eq28}
\epsilon=\Big[{{\hbar}\over{\mu b}}\times
{{N\zeta(D+1)\Gamma({{D+1}\over{2}})}
\over{2^{D+1}\pi^{{3D+3}\over{2}}}}\Big]^{1/D}\  .
\end{eqnarray}
Here we have used the relation $(l_0+b/2)^{-D}\ll (l_0-b/2)^{-D}$,
which is valid in the near-horizon region, $l_0-b/2\ll b$.

From Eq. (\ref{Eq28}) one learns that the neutral point lies in the
near-horizon region (that is, $\epsilon\ll1$) only {\it if} the size
of the box is {\it large} enough such that
\begin{equation}\label{Eq29}
{{b}\over{b_C}}={{N\zeta(D+1)\Gamma({{D+1}\over{2}})}
\over{2^{D+1}\pi^{{3D+3}\over{2}}\epsilon^D}}\  ,
\end{equation}
where $b_C\equiv\hbar/\mu$ is the Compton length of the box. Note
that the RHS of Eq. (\ref{Eq29}) increases very rapidly with the
number $D$ of spatial dimensions: $RHS\sim N
(D/8e\pi^3\epsilon^2)^{D/2}\gg1$ for $D\gg 1$. This implies that for
$D\gg1$, the neutral point can lie in the near-horizon region only
if the box is very large compared to its Compton length. As we shall
now show, this severely restricts our ability to deduce the entropy
bound (\ref{Eq2}) directly from the GSL in $(D+1)$-dimensional
spacetimes with $D\gg1$.

\section{The entropy bound}
The assimilation of the body by the black hole results in a change
$\Delta{\cal M}={\cal E}_{\text{min}}$ in the mass of the black
hole. Using the relation $A_H=A_{D-1}r^{D-1}_H$ for the black hole's
surface area together with Eqs. (\ref{Eq6}) and (\ref{Eq27}), one
finds
\begin{equation}\label{Eq30}
(\Delta A_H)_{\text{min}}=4\pi\mu
b\Big(1+{{2D}\over{D-1}}\epsilon\Big)\
\end{equation}
for the corresponding change in the surface area of the black hole.
[Note that terms of order $(\Delta{\cal M}/{\cal M})^2$ are
negligible for $b\ll r_H$.] Using the entropy-area relation for
black holes, $S_{BH}=A_H/4\hbar$, one finds
\begin{equation}\label{Eq31}
(\Delta S_{BH})_{\text{min}}={{\pi\mu
b}\over{\hbar}}\Big(1+{{2D}\over{D-1}}\epsilon\Big)\
\end{equation}
for the corresponding increase in the black hole's entropy.

Assuming the validity of the GSL [that is, $(\Delta
S)_{\text{tot}}\equiv(\Delta S_{BH})_{\text{min}}-S\ge 0$], one may
deduce an upper bound on the entropy $S$ of a $(D+1)$-dimensional
physical system of proper energy $E$ and proper length $b$:
\begin{equation}\label{Eq32}
S\leq {{\pi bE}\over{\hbar}}\Big(1+{{2D}\over{D-1}}\epsilon\Big)\ ,
\end{equation}
where $\epsilon$ is determined in (\ref{Eq28}).

We see that the second term in (\ref{Eq32}), whose origins are in
the quantum buoyancy, limits our ability to deduce the universal
entropy bound in its canonical form (\ref{Eq2}). So, suppose one is
willing to tolerate a small deviation from the canonical form
(\ref{Eq2}) when deriving the entropy bound from the GSL. For
example, assume that one wants to deduce from the present gedanken
experiment an entropy bound of the form:
\begin{equation}\label{Eq33}
S\leq (1+\delta)\times{{\pi bE}\over{\hbar}}\ ,
\end{equation}
with $\delta\ll1$. This amounts to the requirement that the
magnitude of the work done to overcome the quantum buoyancy, Eq.
(\ref{Eq23}), is limited by $\delta$ times the magnitude of the
energy associated with the body's mass, Eq. (\ref{Eq13}). From
(\ref{Eq32}) one finds that $\epsilon$ should take the small value
$\epsilon=\delta(D-1)/2D$. From Eq. (\ref{Eq28}) with
$\epsilon=\delta(D-1)/2D$ we learn that the entropy bound
(\ref{Eq33}), which is weaker than the canonical bound (\ref{Eq2})
by the factor $1+\delta$, can be deduced from the GSL only for
systems whose size is large enough such that
\begin{equation}\label{Eq34}
b\geq{\hbar\over\mu}\times F(D;\delta)\ ,
\end{equation}
where
\begin{equation}\label{Eq35}
F(D;\delta)\equiv{{N\zeta(D+1)\Gamma({{D+1}\over{2}})D^D}
\over{2\pi^{{3D+3}\over{2}}(D-1)^D\delta^D}}\ \ ;\ \ \delta\ll1\ .
\end{equation}

As an example, let us take $\delta=10^{-1}$ in (\ref{Eq33}). This
amounts to an entropy bound which is $10\%$ weaker than the
canonical form (\ref{Eq2}). For the familiar case of three spatial
dimensions, one then finds a relatively small factor of
$F(D=3)\approx 10^1$, which amounts to the restriction $b\gtrsim
10^1b_C$ on the size of the body. This implies that in three spatial
dimensions, the entropy bound (\ref{Eq33}) with $\delta=10^{-1}$ can
be derived from the GSL even for very small physical systems like
atomic nuclei. However, one soon realizes that the function $F(D)$
in (\ref{Eq35}) increases very rapidly with the number of spatial
dimensions: $F\approx 10^{5},\ 10^{40}$, and $10^{91}$ for $D=10,\
50$, and $100$, respectively. This implies that the entropy bound
(\ref{Eq33}) with $\delta=10^{-1}$ can be deduced from the GSL only
for systems whose size is bounded from below according to: $b\gtrsim
10^5 b_C,\ 10^{40} b_C$, and $10^{91} b_C$, respectively.

Had we taken $\delta=10^{-2}$ in (\ref{Eq33}), we would have found
that an entropy bound which is $1\%$ weaker than the canonical form
(\ref{Eq2}) can be deduced from the GSL only for systems whose size
is large enough such that $b\gtrsim 10^4 b_C,\ 10^{15} b_C,\ 10^{90}
b_C$, and $10^{191} b_C$ for $D=3,\ 10,\ 50,$ and $100$,
respectively.


\section{Fluid vs. wave picture}
Are there any relevant effects which might change our conclusions?
It was pointed out in \cite{Bek4} that the original derivation of
the buoyant force from a fluid picture \cite{UW,Bek5} is valid if
the characteristic wavelength $\bar\lambda$ in the thermal
acceleration radiation is smaller than the box size $b$. At a
fundamental level the buoyant force is caused by the momentum jolts
the box receives as successive waves scatter off it \cite{Bek4}.
Long waves with $\lambda\gg b$ have difficulty matching specified
boundary conditions on the surface of the box. Thus, they tend to
scatter poorly and convey little momentum to the box \cite{Bek4}.
One indeed finds that the wave scattering force is much weaker than
the fluid force \cite{Bek4}.

Note that the thermal distribution $\omega^{D}/(e^{\hbar\omega/T}\mp
1)$ of the acceleration radiation in Eq. (\ref{Eq17}) peaks at the
characteristic frequency
\begin{equation}\label{Eq36}
\bar\omega={{DT}\over{\hbar}}[1\mp e^{-D}+O(e^{-2D})]\  .
\end{equation}
Taking cognizance of Eqs. (\ref{Eq12}) and (\ref{Eq36}), we find
that the characteristic wavelength $\bar\lambda$ of the acceleration
radiation at a proper distance $l$ above the horizon is
\begin{equation}\label{Eq37}
\bar\lambda\simeq {{(2\pi)^2l}\over{D}}\  .
\end{equation}
Thus, one realizes that the fluid picture of the acceleration
radiation which was used in \cite{UW,Bek4,Bek5} (and which requires
$\bar\lambda\lesssim b$) is valid in the region
\begin{equation}\label{Eq38}
l\lesssim l(D)_{\text{fluid}}\equiv{{D}\over{(2\pi)^2}}b\  .
\end{equation}

From Eq. (\ref{Eq38}) one learns that the fluid region actually
extends higher and higher above the black hole as the number $D$ of
spatial dimensions becomes large. This implies, in particular, that
at the near-horizon floating point, the larger is the value of $D$,
the larger is the part of the box which is immersed in the fluid
region. Thus, the larger is the value of $D$, the better is the
fluid description. In fact, from (\ref{Eq38}) we learn that for
$D\gtrsim (2\pi)^2$ the {\it entire} box is already inside the fluid
regime if the floating point is in the near-horizon region,
$l_0-b/2\ll b$. Thus, for $D\gtrsim40$ the use of the fluid picture
in the near-horizon floating point is exact.

For $3\leq D\lesssim 40$ and assuming that the floating point is in
the near-horizon region, part of the body (the lower part) is in the
fluid region (\ref{Eq38}) while part of it is in the (long) wave
scattering regime. For $D$ values in this range, it is therefore
more appropriate to perform the integration in (\ref{Eq22}) in the
range $[l_0-b/2,l(D)_{\text{fluid}}]$, where the fluid picture is
valid:
\begin{equation}\label{Eq39}
W_{\text{buoy}}=A\int_{l_0-b/2}^{l(D)_{\text{fluid}}}
{{N\zeta(D+1)\Gamma({{D+1}\over{2}})(D-2)\hbar}\over{2^{D+2}{\pi}^{{3D+3}\over{2}}r_Hl^{'D}}}dl^{'}\
.
\end{equation}
This work should be corrected for the contribution from the rest of
the box which is in the wave scattering regime. However, we know
that the wave scattering force is much weaker than the fluid force
\cite{Bek4,Notewave}. Hence, the integral in the range
$[l_0-b/2,l(D)_{\text{fluid}}]$ must give a close approximation to
the true work. Indeed, this was verified by direct numerical
computations in \cite{Bek4}. The physical reason for this success of
the fluid picture (provided the floating point is in the
near-horizon region, $l_0-b/2\ll b$) lies in the observation that
the pressure {\it drops} precipitously as $l$ grows, $P\chi\propto
l^{-D}$. This implies that the main contribution to the integrals in
(\ref{Eq22}) and (\ref{Eq39}) comes from the lower part of the box,
which is very close to the horizon well inside the fluid regime.

Performing the integration in (\ref{Eq39}), one finds that for
$3\leq D\lesssim 40$ the expression (\ref{Eq26}) for the energy of
the body should be replaced by
\begin{eqnarray}\label{Eq40}
&&{\cal E}_{\text{min}}={{D-2}\over{2r_H}}\mu \nonumber \\&& \times
\Bigg[l_0+{{b}\over{D-1}}{{\epsilon^{-D+1}-(D/4\pi^2)^{-D+1}}\over{\epsilon^{-D}-(D/4\pi^2)^{-D}}}\Bigg]\
.
\end{eqnarray}
Note that for all $D$ values, $(D/4\pi^2)^{-D}\ll\epsilon^{-D}$ in
the near horizon limit, $\epsilon\ll1$ (as explained above, this
reflects the fact that the pressure drops precipitously as $l$
grows). For example, assume $\epsilon=(D-1)/20D$ which amounts to
$\delta=10^{-1}$ in the bound (\ref{Eq33}). One then finds
$(D/4\pi^2)^{-D}/\epsilon^{-D}\simeq 0.08,\ 10^{-7}$, and $10^{-53}$
for $D=3,\ 10$, and $40$, respectively. Thus, one realizes that Eq.
(\ref{Eq40}) reduces to (\ref{Eq27}) in the near-horizon limit for
all $D$ values.

\section{Summary}
We have considered a gedanken experiment in which a box full of
entropy is lowered towards and then dropped into a $(D +
1)$-dimensional black hole in equilibrium with thermal radiation.
The effects of quantum buoyancy are parameterized by the function
$F(D;\delta)$ defined in Eqs. (\ref{Eq34})-(\ref{Eq35}). We have
seen that an entropy bound of the form (\ref{Eq33}) can be deduced
from the GSL only {\it if} the system is {\it large} enough such
that its size is larger than its Compton length by the factor
$F(D;\delta)$, see Eqs. (\ref{Eq34})-(\ref{Eq35}).

In the familiar case of three spatial dimensions, the factor $F$ is
relatively small for $\delta=O(10^{-1})$. This allows one to deduce
an entropy bound which is $\sim10\%$ weaker than the canonical form
(\ref{Eq2}) directly from the GSL for all macroscopic and mesoscopic
objects (down to the scale of atomic nuclei). However, the function
$F(D;\delta)$ increases very rapidly with the number of spatial
dimension: $F\sim N (D/2\pi^3\delta^2)^{D/2}\gg 1$ for $D\gg 1$. As
a consequence, one finds that quantum buoyancy severely restricts
our ability to deduce the universal entropy bound from the GSL in
higher-dimensional spacetimes with $D\gg1$.

In particular, for physical systems whose size lies in the range
${b/b_C}<F(D;\delta)$, the entropy bound (\ref{Eq2}) cannot be
derived directly from the GSL, not even in its weaker form
(\ref{Eq33}). In other words, for such systems the entropy bound
(\ref{Eq2}) does not serve as a necessary condition for the validity
of the GSL. (Note, however, that the entropy bound (\ref{Eq2}) is
always a sufficient condition for operation of the GSL in this type
of gedanken experiments.) Finally, we note that our findings leave
open the intriguing {\it possibility} of violating the universal
entropy bound (\ref{Eq2}) in the range ${b/b_C}\leq F(D)$, {\it
without} violating the generalized second law of thermodynamics.

\bigskip
\noindent
{\bf ACKNOWLEDGMENTS}

This research is supported by the Meltzer Science Foundation. I
thank Yael Oren and Arbel M. Ongo for helpful discussions. I thank
Jacob D. Bekenstein for helpful correspondence.



\begin{thebibliography}{99}
\bibitem{Notewave} As explained
above, long waves have difficulty matching specified boundary
conditions on the object's surface. As a consequence, they tend to
scatter poorly and convey little momentum to the object \cite{Bek4}.
\end{thebibliography}
\end{document}